\documentclass[11pt,twoside]{article}  
\usepackage{asp2006}
\usepackage{adassconf}

\begin{document}   

%

\paperID{D.17}

%

\title{Software for the Spectral Analysis of Hot Stars}
       
%
%
%
%
%

\markboth{Rauch et al\@.}{Software for the Spectral Analysis of Hot Stars}

%

\author{T\@. Rauch}
\affil{Institute for Astronomy and Astrophysics,
       Kepler Center for Astro and Particle Physics,
       Eberhard Karls University, 
       Sand 1,
       72076 T\"ubingen, 
       Germany}

\author{I\@. Nickelt}
\affil{Astrophysical Institute Potsdam,
       Department of E-Science,
       An der Sternwarte 16, 
       14482 Potsdam, 
       Germany}

\author{U\@. Stampa, M\@. Demleitner}
\affil{Astronomisches Rechen-Institut,
      M\"onchhofstra\ss e 12$-$14,
      69120 Heidelberg, 
      Germany }

\author{L\@. Koesterke}
\affil{Texas Advanced Computer Center, 
       University of Texas, 
       10100 Burnet Road (R8700),
       Austin, 
       Texas 78758$-$4497,
       USA}




\contact{Thomas Rauch}
\email{rauch@astro.uni-tuebingen.de}

%
%

\paindex{Rauch, T.}
\aindex{Nickelt, I.}
\aindex{Stampa, U.}
\aindex{Demleitner, M.}
\aindex{Koesterke, L.}




\setcounter{footnote}{0}


\begin{abstract}          
  In a collaboration of the \emph{German Astrophysical Virtual Observatory} (\emph{GAVO})
and \emph{\mbox{AstroGrid-D}}, the \emph{German Astronomy Community Grid} (\emph{GACG}),
we provide a VO service for the access and the calculation of stellar
synthetic energy distributions (SEDs) based on static as well as expanding 
non-LTE model atmospheres. 

  At three levels, a VO user may directly compare observed and theoretical
SEDs: The easiest and fastest way is to use pre-calculated SEDs from
the \emph{GAVO} database. For individual objects, grids of model atmospheres 
and SEDs can be calculated on the compute resources of \emph{\mbox{AstroGrid-D}} 
within reasonable wallclock time. Experienced VO users may even create 
own atomic-data files for a more detailed analysis.

  This VO service opens also the perspective for a new approach to an 
automated spectral analysis of a large number of observations, e.g. 
provided by multi-object spectrographs.
\end{abstract}


\section{\emph{TheoSSA} -- Model SED on Demand}
\label{sect:TheoSSA}

Spectral analysis by means of Non-LTE model-atmosphere techniques has 
for a long time been regarded as a domain of specialists. Within the 
\emph{German Astrophysical Virtual Observatory} 
(\emph{GAVO}\footnote{http://www.g-vo.org}) project,
we have created the VO service 
\emph{TheoSSA} (Theoretical Simple Spectra Access\footnote{http://vo.ari.uni-heidelberg.de/ssatr-0.01/TrSpectra.jsp}).

\begin{figure}[ht]
\epsscale{1.00}
\plotone{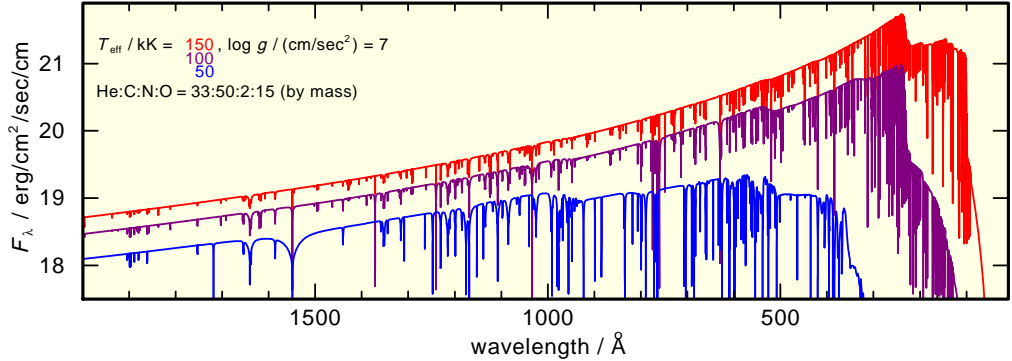}
\caption{SEDs of hydrogen-deficient stars that are available in the \emph{GAVO} database.} 
\label{D.17-fig-1}
\end{figure}

A VO user may 
use pre-calculated grids of spectral energy distributions (SEDs,
in a pilot phase calculated by the \emph{T\"ubingen Model Atmosphere Package}
(\emph{TMAP}\footnote{http://astro.uni-tuebingen.de/\raisebox{.3em}{\tiny$\sim$}rauch/TMAP/TMAP.html})
for hot, compact stars only) which are ready to use and it may be 
interpolated between them to match the user-required parameters. This is 
the easiest way to use synthetic SEDs calculated from Non-LTE model 
atmospheres.

If individual parameters are requested which do not fit to an already 
existing SED in the database, the VO user is guided to
\emph{TMAW}\footnote{http://astro.uni-tuebingen.de/\raisebox{.3em}{\tiny$\sim$}TMAW/TMAW.shtml}.
With this WWW interface, the VO user may calculate an individual model 
atmosphere, requesting effective temperature, surface gravity,
and mass fractions of H, He, C, N, and O (more species will be included 
in the future). For this calculation, standard model atoms are used which 
are provided within the \emph{T\"ubingen Model-Atom Database} 
(\emph{TMAD}\footnote{http://astro.uni-tuebingen.de/\raisebox{.3em}{\tiny$\sim$}rauch/TMAD/TMAD.html}).
Since the VO user can do this without detailed knowledge of the programme 
code working in the background, the access to individually calculated SEDs 
is as simple as the use of pre-calculated SEDs - however,
the calculation needs some time (depending on the number of species 
considered, the wall-clock time is ranging from hours to a few days). 
Standard SEDs (e.g\@. within wavelength ranges of $5-2\,000$\,\AA\ and $3\,000-55\,000$\AA)
of all calculated model atmospheres are automatically ingested 
into the \emph{GAVO} data base and, thus, it is growing in time. Example SEDs are show in Fig.~\ref{D.17-fig-1}.

In case that a detailed spectral analysis is performed, an experienced VO 
user may create an own atomic data file tailored to a specific purpose 
considering all necessary species and calculate own model atmospheres and SEDs.

A similar approach is in preparation, using the newly developed \emph{HotBlast} 
Non-LTE code for spherically expanding atmospheres. \emph{HotBlast} uses as an input 
the atmospheric structure of the static \emph{TMAP} model atmospheres to simulate the 
atmosphere below the wind region.

Spectral analysis by means of NLTE model-atmosphere techniques requires the
calculation of extended grids, varying photospheric parameters like e.g\@.
effective temperature, surface gravity, and element abundances. Since this
is far beyond our computational capacity at the T\"ubingen institute (IAAT), 
calculation of such model-atmosphere grids in the 
framework of \emph{\mbox{AstroGrid-D}} is necessary.

\section{\emph{Atomic Grid Jobs} -- Task Farming with Globus}
\label{theossaatgrid}

\emph{\mbox{AstroGrid-D}}\hspace{.5mm}\footnote{http://www.gac-grid.de/}
is a research and development project that has created a Grid research infrastructure for the 
German astronomical community during the years 2005 to 2009. Based on the Globus Toolkit (GT4) 
middleware, \emph{\mbox{AstroGrid-D}} connects compute cluster and storage resources with desktops 
and specialized hardware, e.g. robotic telescopes (Breitling et al\@. 2008). 
It resulted in several scientific applications and implements novel Grid services, 
such as the information service \emph{Stellaris} (H{\"o}gqvist et al\@. 2007).

Carrying out the \emph{TMAP} model calculations on the compute 
resources of \emph{\mbox{AstroGrid-D}} ensures a much better scaling to a potentially high number
of requests than it would be possible on local compute hardware. Also, the grid as a whole is
more reliable than a local cluster: Even when some of its resources fail,
sufficient alternatives are available, given a stable network connection.

\begin{figure}[t]
\epsscale{1.00}
\plotone{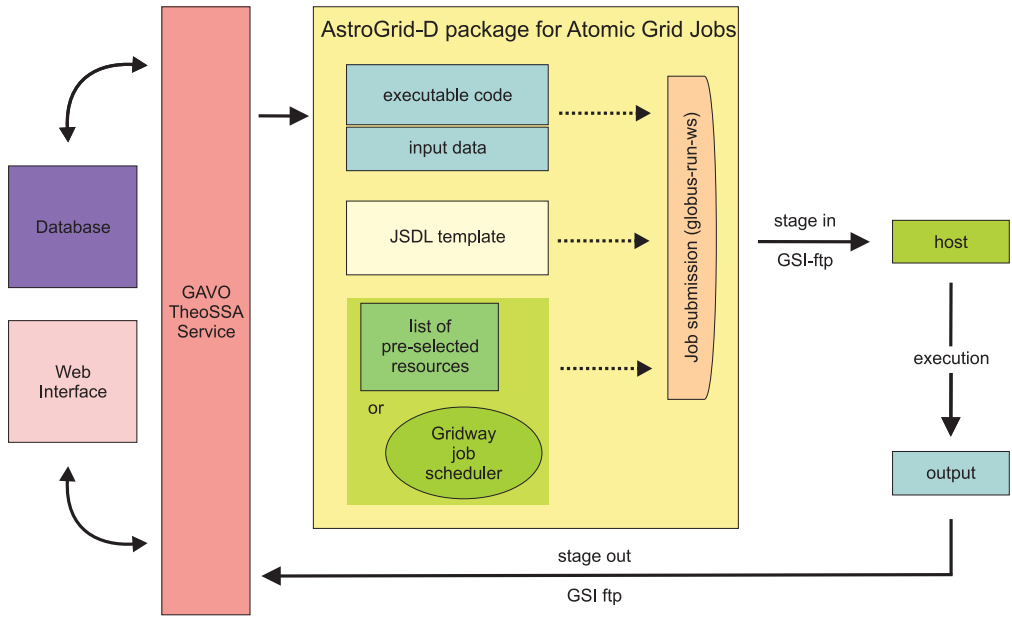}\vspace{10mm}
\caption{Structural diagram: the \emph{TheoSSA} VO service combined with the \emph{Atomic Grid Job package}.} 
\label{D.17-fig-2}
\end{figure}

When a new model is required, the \emph{TheoSSA} service calls a script package that 
independently submits jobs to computational Grid resources. This package was originally 
developed for a different \emph{\mbox{AstroGrid-D}} use 
case\footnote{http://www.gac-grid.org/project-products/Applications/}, but is parameterized 
so that it can easily be adopted to other implementations of so-called \emph{atomic grid jobs}. 
The label \emph{atomic} in this context refers to a specific
type of application for which all input data is transferred together with the software and there 
is no need for inter-process communication. The Grid is thus used as a task farming mechanism. 
The flexible GT4 middleware is able to perform more complex requirements, but we found 
task farming to be an important requirement of many scientific use cases, such as \emph{TMAP}.

Thus transfers of the \emph{TMAP} model-atmosphere software, input data and, upon completion, results 
become part of the standard job submission process in Globus Toolkit. 
The process is further explained in Fig.~\ref{D.17-fig-2}. 
We take advantage of the GT4 web services and control the process by a JSDL template 
(Job Submission Description Language\footnote{http://www.gridforum.org/documents/GFD.56.pdf}). 
Whenever a new job is initiated, the template is 
applied to the specific case and a target machine is selected from a given list. Optionally, a
Grid job broker can be used. Upon completion of the calculation,
the results are passed on to the \emph{TheoSSA} service which stores it in its database and 
notifies the user.

\acknowledgments
We thank the \emph{GAVO} and \emph{\mbox{AstroGrid-D}} teams for support.
The \emph{German Astrophysical Virtual Observatory} (GAVO) project is sponsored by the 
German Federal Ministry of Education and Research (BMBF) under grants 05\,AC6VTB and 05\,AC6VHA.
\emph{\mbox{AstroGrid-D}} ist sponsored by the German Federal Ministry of Education and Research within 
the \emph{D-Grid initiative} under contracts 01\,AK804[A-G].

\end{document}